\documentclass{aa}  
\usepackage{graphicx}
\usepackage{txfonts}
\usepackage{longtable}
\usepackage{supertabular,booktabs}
\newcommand{\mygi}{MyGIsFOS}

\newcommand{\loggf}{\ensuremath{\log\,gf}}
\newcommand{\logg}{\ensuremath{\log\,g}}
\def\teff{$T\rm_{eff}$}
\newcommand{\kms}{$\rm km s ^{-1}$}
\newcommand{\cobold}{CO$^{5}$BOLD}
\newcommand{\leol}{SDSS\,J102915.14+172927.9}
\newcommand{\leo}{SDSS\,J102915+172927}

\begin{document} 

\title{SDSS J102915.14+172927.9: Revisiting the chemical pattern
\thanks{Based on observations made with UVES at VLT 286.D-5045 and 112.25F3.001.}
}
\titlerunning{SDSS J102915.14+172927.9: Revisiting the chemical pattern}

\author{
E.~Caffau    \inst{1,2} \and
P.~Bonifacio \inst{1,2} \and
L.~Monaco \inst{3,2} \and
M.~Steffen \inst{4} \and
L.~Sbordone \inst{5} \and      
M.~Spite \inst{1} \and
P.~Fran\c{c}ois \inst{6,7} \and
A.J.~Gallagher \inst{4} \and
H.-G.~Ludwig \inst{8} \and
P.~Molaro \inst{2,9} 
}

\institute{GEPI, Observatoire de Paris, Universit\'{e} PSL, CNRS,  5 Place Jules Janssen, 92190 Meudon, France
\and
INAF-Osservatorio  Astronomico  di  Trieste,  Via  G.B.  Tiepolo  11,34143 Trieste, Italy
\and
Universidad Andres Bello, Facultad de Ciencias Exactas, Departamento de Ciencias F{\'\i}sicas - Instituto de Astrof{\'\i}sica, Autopista
Concepci{\'o}n-Talcahuano, 7100, Talcahuano, Chile
\and
Leibniz-Institut f\"ur Astrophysik Potsdam, An der Sternwarte 16, 14482 Potsdam, Germany
\and
European Southern Observatory, Alonso de Cordova 3107, Vitacura, Santiago, Chile
\and
GEPI, Observatoire de Paris, Universit\'{e} PSL, CNRS, 77 Av. Dendert-Rocheau, 75014 Paris, France
\and
UPJV, Universit\'e de Picardie Jules Verne, 33 rue St Leu, 80080 Amiens, France
\and
Landessternwarte - Zentrum f\"ur Astronomie der Universit\"at Heidelberg, K\"onigstuhl 12, 69117 Heidelberg, Germany
\and
Institute  of  Fundamental  Physics  of  the  Universe,  Via  Beirut  2,34151 Trieste, Italy
}

   \date{Received July 15, 2024; accepted August 16, 2024}

  \abstract
{The small- to intermediate-mass ($\rm M<0.8 M_\odot$), most metal-poor stars that formed in the infancy of the Universe are still shining today in the sky. They are very rare, but their discovery and investigation brings new knowledge on the formation of the first stellar generations.
}
{SDSS J102915.14+172927.9 is one of the most metal-poor star known to date. Since no carbon can be detected in its spectrum, a careful upper limit is important, both to classify this star and to distinguish it from the carbon-enhanced stars that represent the majority at these metallicities.
}
{We undertook a new observational campaign to acquire high-resolution UVES spectra.
The new spectra were combined with archival spectra in order to increase the signal-to-noise ratio. 
From the combined spectrum, we derived abundances for seven elements (Mg, Si, Ca, Ti, Fe, Ni, and a tentative Li) and five significant upper limits (C, Na, Al, Sr, and Ba).
}
{The star has a carbon abundance $\rm A(C)<4.68$ and therefore is not enhanced in carbon, at variance with the majority of the stars at this Fe regime, which typically show $\rm A(C)>6.0$. A feature compatible with the Li doublet at 670.7\,nm is tentatively detected. 
}
{The upper limit on carbon implies  $\rm Z<1.915\times 10^{-6}$, more than 20 times lower than the most iron-poor star known. Therefore, the gas cloud out of which the star was formed did not cool 
via atomic lines but probably through dust.
Fragmentation of the primordial cloud is another possibility for the formation of a star with a metallicity this low.}

   \maketitle
  \nolinenumbers
%
\section{Introduction\label{intro}}

The process of formation of the first generation of stars (Pop\,III), formed after the Big Bang, is still not clearly understood and remains to be identified \citep[see][]{bromm13}. One of the main questions is whether low-mass stars can form from the primordial material, composed of isotopes of H, He, and traces of Li. 
A star of mass $M\approx0.8\,{\rm M}_\odot$ or lower that formed in the first 100 Myr after the Big Bang would still
be shining today on the main sequence. The fact that, to date, such a star has not been observed
could serve as an argument against the formation of such primordial stars, yet the absence of evidence is not
evidence of absence. 
Theoretically, the difficulty of forming a low-mass star from primordial
matter is due to the lack of an efficient cooling mechanism that allows the star-forming
gas cloud to continue to collapse instead of being stopped by the increasing radiative pressure caused by
the consequential heating from compression. A star of large mass can be formed, since the gas cloud's gravity
is strong enough to overcome the pressure resulting from this heating \citep[see ][and references therein]{klessen23}.

\citet{greif10}  showed that fragmentation of a large gas cloud can even take place
for a primordial chemical composition resulting in the formation of low-mass primordial stars.
However, the simulation could not be run long enough to be sure that the fragments will
not merge again giving rise to a single massive star.
\citet{bromm03} showed that if there is a small amount of C and O in the gas,
the cooling goes through collisional excitation and radiative de-excitation. 
This led \citet{frebel07} to introduce the transition discriminant
$\rm D=\log\left [ 10^{[C/H]}+0.3\times10^{[O/H]}\right ]$. The stars
formed through atomic line cooling are characterised by $\rm D \ge D_{crit}\simeq -3.5\pm 0.2$.
The presence of carbon and oxygen in a primordial gas cloud provides a means of cooling the contracting gas to allow the formation
of low-mass stars. But there is another way to cool at a chemical composition close to the pristine one.
\citet{schneider12} proposed dust as a a cooling mechanism for the formation of a second generation of low-mass stars.
\citet{smith15} investigated the formation of the transition between Pop\,III and Pop\,II stars, suggesting that
multiple scenarios probably contributed to this transition.

The star \leol\ (also known as {\it Gaia}\,DR3\,3890626773968983296, hereafter \leo) 
was first studied by \citet{leo11} who claimed,
based on UVES spectra, a very low iron abundance, $\rm [Fe/H]=-4.73$, and a very low carbon abundance upper limit, $\rm A(C) \le 4.7$. 
These authors found that the transition discriminant in this star is below the critical value defined above,
thus invoking the cooling through dust \citep{omukai05} to explain the formation of this star.
The formation of \leo\ was theoretically investigated by several authors, and the conclusion was
that the cooling was due to dust \citep[see][]{schneider12,klessen12,chiaki14,bovino16}.
\citet{leo11} claimed that this star was the most metal-poor object known in the Universe, with $\rm Z< 1.5\times 10^{-5}$, at the time of writing.
\citet{hattori14} later suggested that \leo\ was formed from a primordial gas by zero-metal cooling, that is only
through Ly$\alpha$ and $\rm H_2$ cooling \citep{Silk77}, and was later chemically enriched by accreting supernovae ejecta. 

Clearly at the time of the discovery of \leo, there was no way to derive an accurate surface gravity; the photometry
available was consistent not only for a star on the main sequence, but also for a turn-off or sub-giant branch star.
\citet{leo11} adopted an intermediate gravity of 4.0\,dex, corresponding to the turn-off solution.
\citet{macdonald13} claimed that the star is more likely to be a subgiant star.
Their explanation for the low metal abundance of this star was that
it is the result of 'gravitational settling on the main sequence followed by incomplete convective dredge-up during subgiant evolution'.
In this scenario, the fact that Fe is enhanced with respect to C
implied that it was formed in an environment enriched by Type Ia supernova ejecta rich in Fe. 
With the advent of {\it Gaia}, the parallax put the star on the main sequence \citep[see][]{leo18}. Furthermore, the investigation of Ca ionisation conducted by \citet{sitnova19},
which took into account the effects due to departures from the local thermodynamical equilibrium (LTE), that is so-called non-LTE (NLTE) effects.
This ruled out the \citet{macdonald13} scenario.

Recently, a more sophisticated analysis of the same UVES data used in \citet{leo11}, 
by \citet{lagae23}, using a 3D stellar model and a NLTE treatment of the line formation for some elements, 
claimed that these new assumptions implied that the star was not as iron poor
as was suggested in \citet{leo12} in 1D-NLTE.
Furthermore, \citet{lagae23} derived a much higher upper limit
on C ($\rm A(C)< 5.39$, the upper limit assuming a 1D model and local thermodynamical equilibrium (1D-LTE) 
is added with the uncertainty assigned by the authors), making the star's transition discriminant compatible with its
formation through atomic line cooling.

Since the carbon abundance determination is crucial to understanding whether the star
formed through line cooling or dust cooling,
we analysed newly observed high-resolution spectra that we combined with the previous observations. This
allowed us to improve the signal-to-noise ratio (S/N) and to improve our estimates of the upper limits of several elements,
including carbon.
 
\section{Observations} 
\begin{figure*}
\centering
\includegraphics[width=\hsize,clip=true]{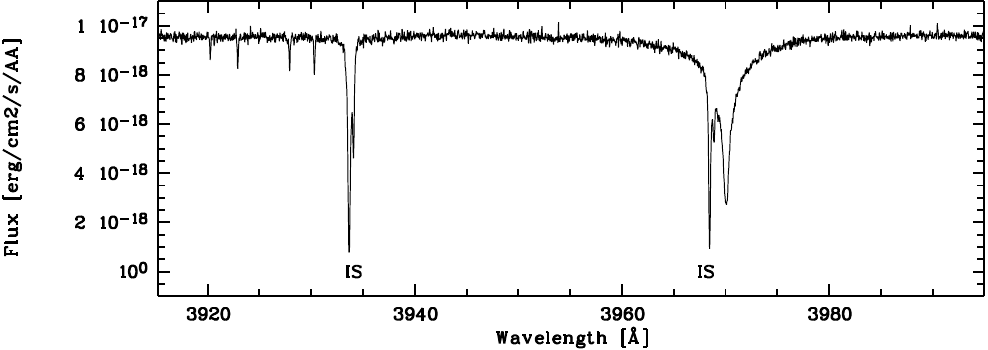}
\caption{Combined observed spectrum flux calibrated in the wavelength range of \ion{Ca}{ii}-K and -H. The two interstellar components are labelled as IS.
The H line and few metallic lines (\ion{Fe}{i}) are visible.}
\label{fig:obs}
\end{figure*}

For this investigation, we used the UVES spectra observed in the DDT programme 286.D-5045 of 2011 (7\,h observations)
and the specifically requested spectra observed during programme 112.25F3.001 (22\,h observations). 
In programme 112.25F3.001, we observed  
in the setting DIC2\,437+760 (wavelength ranges 373--499 and 565--946\,nm),
with slit 1\farcs{4}\ (resolving power of 40\,000).
We reduced all the spectra with the ESO pipeline (on gasgano version), the option of absolute flux calibration and, for the blue-arm spectra, we used the option `noappend' in such a way as to keep all the orders separated.
We corrected all the spectra for the barycentric velocity and derived the radial velocity for each observation. We shifted each spectrum for its radial velocity and merged the orders, as done for the ESPaDOnS spectra in \citet{lombardo21}.
We averaged all the spectra, and the spectrum obtained was used to derive the abundances (see Fig.\,\ref{fig:obs}).

A crucial point in the analysis of the G-band to derive the carbon abundance is the order merging of the UVES spectra.
In the data reduction, we flux-calibrated the spectra, using the instrument response function derived from the
observation of spectrophotometric standard stars\footnote{see \url{https://ftp.eso.org/pub/dfs/pipelines/instruments/uves/uves-pipeline-manual-6.4.1.pdf} section 11.1.20}.
As a test, from the UVES spectra we extracted the two orders where the CH molecules are the strongest in the G-band,
corresponding to wavelength range 429--433\,nm, that is order 108 and 109.
We added all the spectra from prog. ID 286.D-5045 and 112.25F3.001, separately. 
We averaged the two added spectra.
We then used the {\it Gaia}-XP spectrum as the flux for the star, by scaling the UVES spectra on the {\it Gaia}-XP \citep{montegriffo23}.
This step corrects for any flux lost at the slit during the observation.
We  merged the two spectra extracted from the two orders.
We rebinned the {\it Gaia}-XP spectrum on the UVES spectra and then divided the two.
The obtained spectrum is the normalised one that we compared to the complete merge spectrum, selected and normalised in the G-band wavelength range.
The two spectra are indiscernible, so we decided to use the complete merged spectrum for all the chemical investigations.
We underline that the overlap between order 109 and 108 is in the range 429.5 to 430.5\,nm,
which is where there are some of the strongest CH lines.

\section{Model atmospheres}

We computed an ATLAS\,12 model \citep{k05} with the stellar parameters discussed in Sec.\,\ref{sec:param} and abundances derived by \citet{leo12}.
This model has been used to derive the chemical abundances and the upper limits in the present work.

\begin{figure}
\centering
\includegraphics[width=\hsize,clip=true]{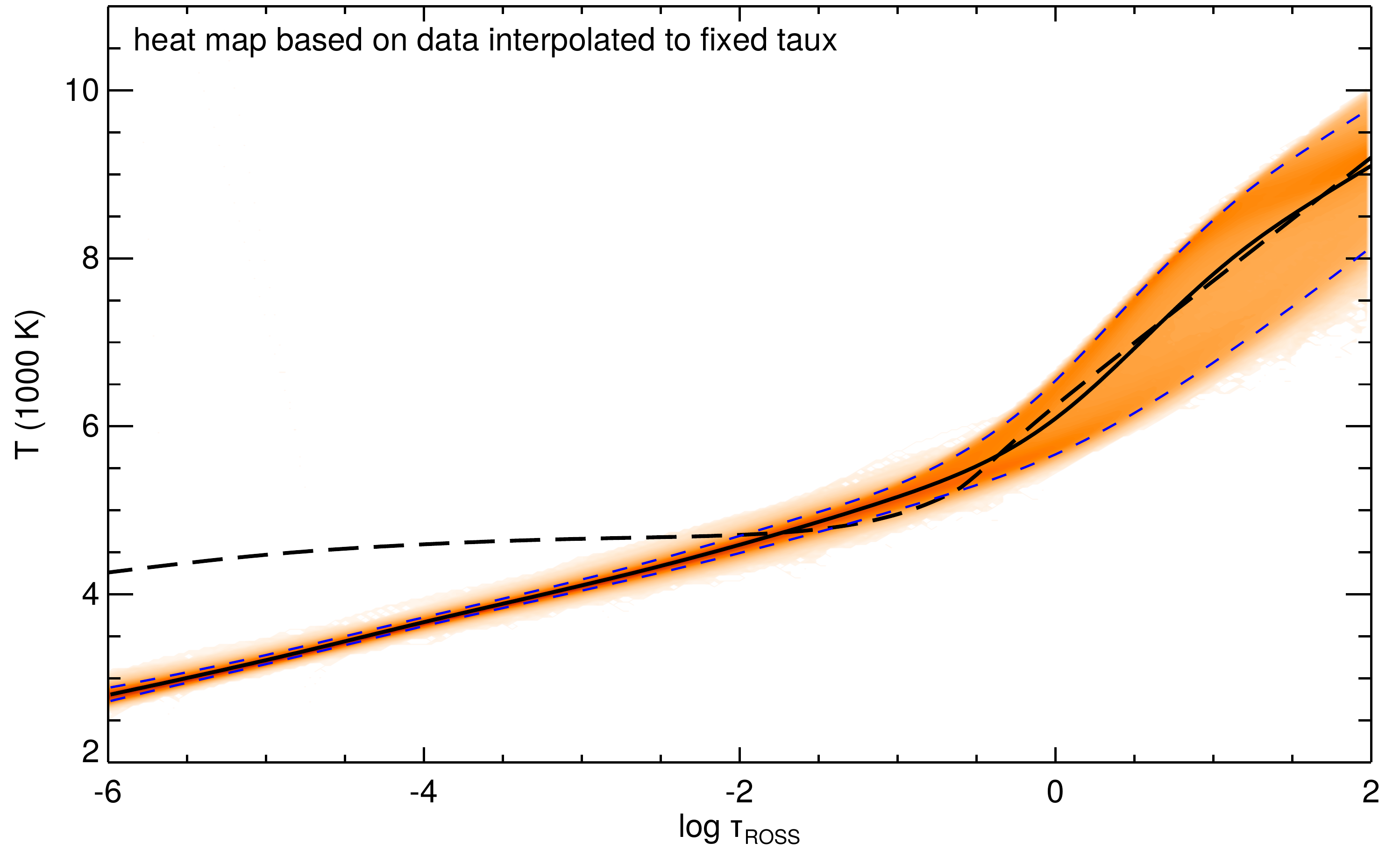}\\
\caption{Three-dimensional versus 1D temperature structure for stellar parameters \teff=$5773$\,K,
\logg=$4.7$, and metallicity [M/H]=$-4.0$. The orange band outlines the
3D temperature distribution of the high-resolution model resampled to $70\times 70\times 160$ grid points. The width of the temperature distribution encountered on surfaces of equal Rosseland optical depth is indicated by the dashed blue lines enclosing 95,5\% of the data points at each height. The solid black line shows the average temperature of the 3D model, the long-dashed black line represents the $T(\tau_{\rm Ross})$ relation predicted by a 1D model atmosphere with identical stellar parameters, treating convection by standard mixing-length theory, but otherwise using the same input physics as the 3D hydrodynamical model.}
\label{fig:ttau}
\end{figure}

We also computed two dedicated 3D hydrodynamical model atmospheres with the latest
version of the \cobold\ code \citep[see][plus updates]{freytag2012} for
stellar parameters that closely represent this star (\teff=$5773$\,K,
\logg=$4.7$, [M/H]=$-4.0$, see Sec.\,\ref{sec:param}). The low-resolution model
has $140\times 140\times 160$ grid cells, and the high-resolution models covers
the same volume by $280\times 280\times 160$ cells. The frequency dependence
of the radiative opacity is represented by 11 bins, each comprising opacities
of similar strength. Figure\,\ref{fig:ttau} depicts the temperature stratification of
the resampled high-resolution model, which is almost identical to that of its
low-resolution counterpart. 

As is well known, the temperature stratification of the photospheres of
metal-poor stars is sensitive to hydrodynamical flows that cannot be modelled
in 1D. This implies a systematic photospheric temperature difference between
standard 1D model atmospheres where convection is treated by the mixing-length
theory and 3D hydrodynamical simulations where convective overshoot is
emerging naturally. For the stellar parameters considered here, the
temperature of the 3D model is cooler than predicted by the corresponding 1D
model for optical depths $\log \tau_{\rm Ross} < -2.0$, whereas the
temperature difference is reversed in the range $-2.0 < \log \tau_{\rm Ross} <
-0.5$ (compare solid and dashed lines in Fig.\,\ref{fig:ttau}). It is
noteworthy that the amplitude of the temperature fluctuations in the upper
photosphere is remarkably small; the 3D model can probably be represented
reasonably well by its <3D> mean structure. However, we make use of the full
3D results in this work. The low amplitude of the photospheric temperature fluctuations was already recognised in the 3D model for the star used by \citet{leo12} and tentatively traced back to the damping influence of an increased specific heat caused by H$_2$ molecular formation. \citet[][see their Fig.\,1]{thygesen2017} show that small temperature fluctuations are a common property of metal-poor 3D models in the considered regime of effective temperatures and surface gravity.

As mentioned above, all chemical abundance determinations in this work
(including the upper limit estimates) are based on ATLAS\,12 state-of-the-art
1D model atmospheres together with the line formation code SYNTHE
\citep{k05}, used to compute synthetic spectra from the ATLAS\,12 models.
In the case of carbon only, we consider the 3D correction described in Sect.\,\ref{sec:carbon}, resulting in a significantly lower C abundance.

\section{Analysis}

\subsection{Radial velocity} \label{sec:vr}

We derived the radial velocity from all the spectra, already corrected for the barycentric velocity with template matching.
We compared each observed spectrum with a synthetic one.
We find $\rm \langle V_r\rangle =-34.9\pm 0.6$\,\kms, which is compatible with no radial velocity variation from the UVES spectra.
We shifted each spectrum for the derived radial velocity before adding them, because slightly different radial velocity values can be due to the centring of the star on the slit.

\subsection{Stellar parameters} \label{sec:param}

We derived the stellar parameters using the {\it Gaia}\,DR3 \citep{gaiadr3} photometry and parallax.
As in \citet{lombardo21}, we compared the $G_{BP}-G_{RP}$ {\it Gaia} colour to a grid of synthetic colours to derive the effective temperature, \teff.
Once the \teff\ was obtained, we derived the gravity from the {\it Gaia} parallax, corrected by the zero-point as suggested by \citet{lindegren21}, using the Stefan-Boltzmann equation.
For the mass, we compared the stellar parameters to MIST \citep[][]{dotter16,choi16}, BASTI \citep[][]{hidalgo18,pietrinferni21}, and Chieffi FRANEC isochrones (Chieffi private communication) 
and converged to $\rm 0.65 M_\odot$.
This is in perfect agreement with the masses ($\rm 0.62 M_\odot$ to $0.70 M_\odot$) reported in Table\,2 by \citet{leo12}.
If the corrections were not applied to the zero-point of the parallax, the gravity would become 0.05\,dex higher and the temperature 6\,K cooler. This would lead to negligible changes in the abundances. 

The extinction we adopted was $\rm A_V =0.08$ (Rosine Lallement private communication), and at each step we interpolated the extinction coefficient in a grid of synthetic coefficients.
Assuming a metallicity of $-4.7$ for the first iteration, we derived the initial estimate of the stellar parameters and from them the metallicity. 
The final stellar parameters we adopted to compute the ATLAS\,12 model were: \teff=5780\,K, \logg=4.60, no $\alpha$-enhancement, and a microturbulence ($\xi$) of 1\,\kms.
The conclusion on the absence of $\alpha$-enhancement is based on Mg and Ca. The microturbulence is assumed, as there are not enough \ion{Fe}{i} lines with various line strengths to derive it. 

{\bf
}

The \ion{Fe}{i} lines used to derive the metallicity are all in a limited lower energy range ($\rm E_{low}<1.61$\,eV), making them not useful
for deriving \teff\ from a null-trend of A(Fe) with $\rm E_{low}$.
If we group the \ion{Fe}{i} lines in three sets of energy range, we obtain $\rm [Fe/H]=-4.64$ from lines in the range $0-0.85$\,eV, $\rm [Fe/H]=-4.77$ for lines in the range $0.85-1.02$\,eV, and $\rm [Fe/H]=-4.78$ for lines in the range $1.4-1.61$\,eV.
Among the two sets of highest $\rm E_{low}$, there is no trend, as in \citet{cayrel04}, while the lines in the set with the lowest energy provide,
on average, a higher A(Fe) \citep[as in][]{cayrel04}.

\subsection{Kinematics} \label{sec:kine}

In order to evaluate the kinematic properties of \leo, 
we used the same techniques presented in \citet{bonifacio24} and \citet{caffau24}. 
In particular, we used the {\it Gaia}\,DR3 \citep{gaiadr3} coordinates, parallax, and proper motions, and the radial velocity 
we measured from our spectra, as input to the Galpy code \citep[][]{bovy15}. The parallax was corrected for the zero-point 
following \citet[][]{lindegren21}. We adopted the standard Milky Way potential MWPotential2014, 
the solar peculiar motions by \citet[][]{schonrich10}, a distance of the Sun from the Galactic centre of 8\,kpc, and a circular velocity 
at the solar distance of 220\,\kms\ \citep[][]{bovy12}. The stellar orbit was back-integrated for 1\,Gyr. 
We estimated the errors on the calculated quantities as the standard deviation of the values obtained, repeating 
the calculations with a 1000 random realisations of the input parameters. We used the pyia code \citep[][]{pyia} 
to perform the extractions from a multivariate Gaussian, which adopts the errors in the input parameters as standard 
deviations and takes into account the correlation between the input parameters.

\begin{figure}
\centering
\includegraphics[width=\hsize,clip=true]{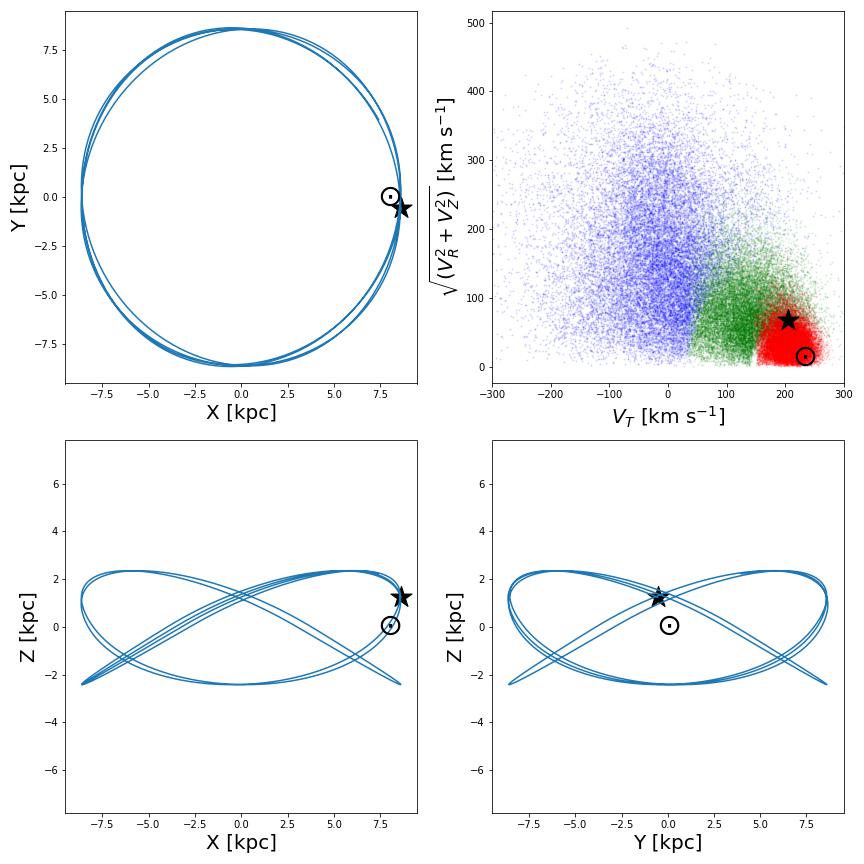}\\
\caption{
Orbit of SDSS J102915+172927 in the Y vs X (upper-left), Z vs X (bottom-left), and Z vs Y (bottom-right) planes.  X, Y, and Z are Galactocentric Cartesian coordinates. Upper-right panel: $\sqrt(V_R^2+V_Z^2)$ vs $\rm V_T$ (Toomre diagram: Transversal velocity component vs a combination of the radial and vertical velocity components in Galactocentric cylindrical coordinates). The coloured points are the stars of the  `'good parallax sub-sample' of \citet[][]{topos6}, plotted for reference. They are divided into thin (red) disc, thick (green) disc, and halo (blue) stars, following the \citet[][]{bensby14} criteria. The filled star marks SDSS J102915+172927, while the Sun is indicated by the $\odot$ symbol.}
\label{fig:kine}
\end{figure}

Figure \ref{fig:kine} shows the \leo\ orbit in X, Y, and Z rectangular Galactocentric coordinates (bottom-left, bottom-right and upper-left panels). The current position of the star is indicated as a black solid star. \leo\ lies at a distance of 1.47$\pm0.13$\,kpc from the Sun and of 8.72$\pm 0.07$\,kpc from the Galactic centre, $1.24\pm 0.11$\,kpc above the Galactic plane. Its orbit is pro-grade, confined within  $\rm Z_{\rm max}=2.42\pm0.26$\,kpc of the Galactic plane and of low eccentricity ($e=0.02\pm 0.01$), with peri- and apo-Galactocentric distances of $\rm r_{\rm peri}=8.64\pm 0.05$\,kpc and $\rm r_{\rm ap}=8.93\pm 0.09$\,kpc. The upper-right panel shows the \leo\ position in the Toomre diagram, namely the transversal velocity component, versus a combination of the radial and vertical velocity components in Galactocentric cylindrical coordinates. We plotted for reference, the  `good parallax sub-sample' of \citet[][]{topos6}, where stars are classified as belonging to the thin disc (red), thick disc (green), and halo (blue), according to the scheme adopted by \citet[][]{bensby14}. \leo\ is classified as a `thick disc' star according to this scheme. 

The kinematics of \leo\ have also been previously investigated by \citet[][]{sestito19}, \citet[][]{dimatteo20} and \citet[][]{dovgal24}. In particular, \citet[][]{dovgal24} followed the same methodology as \citet[][]{sestito19} but
used {\it Gaia}\,DR3 astrometric parameters. \citet[][]{dimatteo20} inverted 
the {\it Gaia}\,DR2 parallax corrected by a zero-point of $-0.03$ to derive the distance (1.31\,kpc), while \citet[][]{dovgal24} inferred a distance from the Sun of $1.49\pm 0.32$\,kpc.

Our results are in good agreement with both analyses. \citet[][]{dovgal24} indicate that the star is on a pro-grade, almost circular ($e=0.09\pm0.02$) disc orbit around the Galactic centre, confined to 
$\rm Z_{\rm max}=2.36\pm0.60$\,kpc of the Galactic plane. Similarly to us, these authors used the galpy code 
and the MWPotential2014 potential, which, however, they modified to have a more massive dark matter halo. 
\citet[][]{dimatteo20}, who used different codes and Galactic potential, also obtain for this star a quasi-circular 
orbit ($e=0.04,$ $\rm r_{\rm peri}=8.9\pm0.0$\,kpc and $\rm r_{\rm ap}=9.6\pm0.2$\,kpc), confined close to the Galactic plane ($\rm Z_{\rm max}=2.2$\,kpc).

\subsection{Abundances}\label{secabbo}

To derive the abundances, we used \mygi\ \citep{mygi14} but in a non-standard way. 
We computed an ATLAS\,12 model \citep{k05} with the abundances derived by \citet{leo12}.
With SYNTHE \citep{k05}, using the ATLAS\,12 model (thus, keeping the same temperature structure), 
we computed a grid of synthetic spectra varying in steps of 0.2\,dex in abundance for all elements.
We built a grid with these syntheses, to be used by \mygi.
We selected the ranges with no lines to be used to pseudo-normalise the observed spectrum and in the same way the synthetic spectra.
We verified that the ranges to be used to pseudo-normalise the spectra were clean from unexpected emission or absorption 
(e.g. telluric lines), and we verified all fits. 
We derived abundances for six elements (see Table\,\ref{tab:abbo}).
As expected, the abundances derived are in agreement with \citet[][]{leo11,leo12}.
Due to the high quality of the spectrum, the Fe abundance is now based on 56 \ion{Fe}{i} features, and not 43 as was the case in \citet{leo12}. We derived $\rm [Fe/H]=-4.73\pm 0.09$\,dex.
For the 1D-NLTE correction on iron, \citet{leo12} derived $+0.13$\,dex from three \ion{Fe}{i} lines. This value is in agreement with $+0.12$\,dex derived 
from 32 \ion{Fe}{i} lines when using the corrections from \citet{bergemann12fe}. 
The differences with other 1D-NLTE corrections for \ion{Fe}{i} in this star found
in the literature are rooted in the different assumptions made to take
into account the collisions with hydrogen atoms.
The 1D-NLTE correction of $+0.25$\,dex derived by \citet{lagae23} is 
in agreement with 
the NLTE correction of $+0.24$\,dex \citep{leo12} adopting $\rm S_H=0.1$.
\citet{ezzeddine17} derived a NLTE correction for Fe of +0.4\,dex using their semi-empirical
quantum fitting method \citep{ezzeddine2018} to estimate the effect of collisions
with hydrogen atoms.
We adopt the value computed by \citet{leo12} for the NLTE correction on Fe with $\rm S_H=1$, a choice that was carefully taken at the time.
In addition, we favour a smaller NLTE correction in Fe because there are some extremely metal-poor stars that are not enhanced or slightly poor in $\alpha$ elements \citep[see e.g.][]{caffau13}, but very few are poor in $\alpha$ elements \citep[see][]{li22}. 
Increasing the Fe abundance would make this star $\alpha-$poor.

Here we did not use the \ion{Mg}{i}-b lines used by \citet{leo12} because only the 2011 observations cover this wavelength range and the S/N is low. Instead, we used only the UV lines that with combined spectra provide a high S/N of 78.
From the three \ion{Mg}{i} lines (382.9, 383.2 and 383.8\,nm), we derived $\rm [Mg/H]=-4.67\pm 0.04$.
The agreement with \citet{leo12} is good ($\rm [Mg/H]=-4.73$ from these three lines).
For the \ion{Mg}{i} lines used here, \citet{leo12} has a NLTE correction of $+0.13$\,dex, which is in agreement with $+0.17$ derived by using the corrections
from \citet{bergemann17} and with the value in \citet{lagae23}.

For the \ion{Si}{i} line at 390.5\,nm, already investigated in \citet{leo12}, we derived a lower Si abundance ($\rm [Si/H]=-4.37$), but we have a better quality spectrum (S/N=84).
The NLTE correction in \citet{leo12} is $+0.35$\,dex for the dwarf solution, the value that we adopt here.
From the 422\,nm \ion{Ca}{i} line, we derived $\rm [Ca/H]=-4.76$. A NLTE correction of $+0.24$ was derived by \citet{leo12} for the dwarf solution.
For Ti we investigated only the \ion{Ti}{ii} line at 336.1\,nm. According to \citet{sitnova_ti}, Ti ionised lines form close to the LTE condition.
We kept five \ion{Ni}{i} lines and derived $\rm [Ni/H]=-4.64\pm 0.15$.

\begin{table*}
\caption{Abundances derived for \leo. The solar abundances are from \citet{caffausun} and \citet{lodders09}.}
\label{tab:abbo}
\begin{tabular}{lrlrrrrrrrrr}
\hline
\smallskip
X & Z  &ion & $\rm A(X)_\odot$ &Nlines& A(X)     & [X/H]      & $\sigma$  & [X/Fe] &   NLTE cor. & A(X)L & A(X)C \\ 
\hline
\smallskip
C  &  6 &  0 & 8.50 & G-band &$ <4.68 $ & $<-3.82 $  &           & $< 0.91 $ &         & $<5.25$ & $< 4.70$ \\ 
Li &  3 &  0 &      &        &$  1.08:$ &            &           &           &         & $<1.02$ &          \\ 
Na & 11 &  0 & 6.30 &        &$ <1.23 $ & $ -5.07 $  &           & $<-0.34 $ &         & $<1.55$ &          \\ 
Mg & 12 &  0 & 7.54 &  3     &$  2.87 $ & $ -4.67 $  & 0.04      & $  0.06 $ & $+0.13$ & $ 2.90$ & $  2.83$ \\
Al & 13 &  0 & 6.47 &        &$ <1.23 $ & $<-5.24 $  &           & $<-0.51 $ &         & $<1.51$ &          \\
Si & 14 &  0 & 7.52 &  1     &$  3.15 $ & $ -4.37 $  & 0.15      & $  0.36 $ & $+0.35$ & $ 3.16$ & $  3.25$ \\
Ca & 20 &  0 & 6.33 &  1     &$  1.57 $ & $ -4.76 $  & 0.15      & $ -0.03 $ & $+0.24$ & $ 1.58$ & $  1.61$ \\
Ti & 22 &  1 & 4.90 &  1     &$  0.22 $ & $ -4.68 $  &           & $  0.05 $ &         & $ 0.54$ & $  0.15$ \\
Fe & 26 &  0 & 7.52 & 56     &$  2.79 $ & $ -4.73 $  & 0.09      & $  0.00 $ & $+0.12$ & $ 2.80$ & $  2.79$ \\
Ni & 28 &  0 & 6.23 &  5     &$  1.59 $ & $ -4.64 $  & 0.15      & $  0.09 $ &         & $ 1.71$ & $  1.68$ \\
Sr & 38 &  1 & 2.92 &        &$<-2.19 $ & $<-5.11 $  &           & $<-0.38 $ & $+0.14$ &         & $<-2.18$ \\
Ba & 56 &  1 & 2.17 &        &$<-2.09 $ & $<-4.26 $  &           & $< 0.47 $ &         &         &          \\
\hline
\end{tabular}
\\
Notes:
{In column 11 (A(X)L) the 1D-LTE abundances from \citet{lagae23} derived with similar stellar parameters (\teff=5811, \logg=4.68),
and in column 12 (A(X)C) the abundances from \citet{leo12}, with a similar \teff\ of 5811 and a much lower \logg\ of 4.0, are given for reference.}
\end{table*}

\begin{figure}
\centering
\includegraphics[width=\hsize,clip=true]{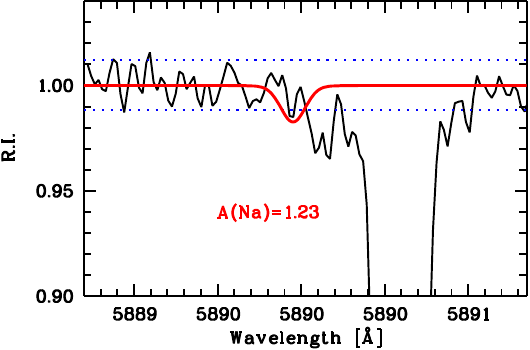}
\caption{Observed spectrum (solid black) in the wavelength of the \ion{Na}{i} D1 line
compared to a synthesis (solid red) with A(Na) derived from Cayrel's formula multiplied by a factor 3.
The S/N of 85 is highlighted by the dashed blue lines.}
\label{fig:naul}
\end{figure}

\begin{figure}
\centering
\includegraphics[width=\hsize,clip=true]{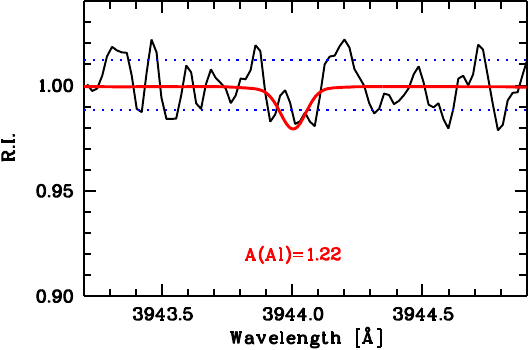}
\caption{Observed spectrum (solid black) in the wavelength of the 349.9\,nm \ion{Al}{i} line
compared to a synthesis (solid red) with A(Al) from the upper limit.
The S/N of 85 is highlighted by the dashed blue lines.}
\label{fig:alul}
\end{figure}

\begin{figure}
\centering
\includegraphics[width=\hsize,clip=true]{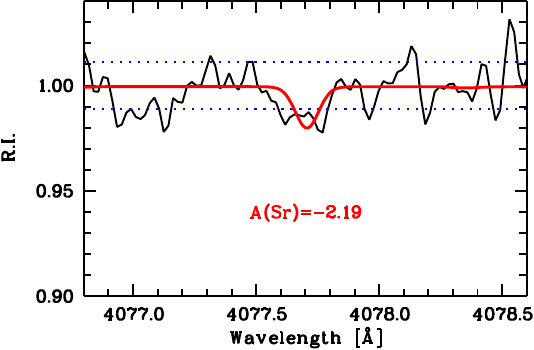}
\caption{Observed spectrum (solid black) in the wavelength of the 407.7\,nm \ion{Sr}{ii} line
compared to a synthesis (solid red) with A(Sr) from the upper limit.
The S/N of 90 ratio is highlighted by the dashed blue lines.}
\label{fig:srul}
\end{figure}

\begin{figure}
\centering
\includegraphics[width=\hsize,clip=true]{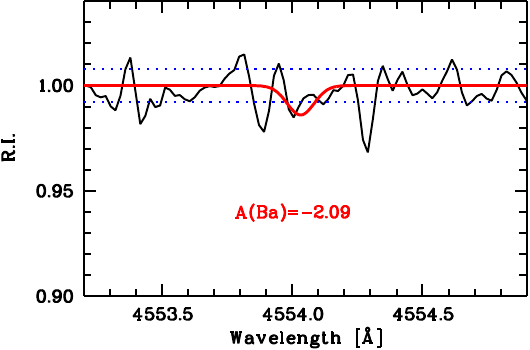}
\caption{Observed spectrum (solid black) in the wavelength of the 455.4\,nm \ion{Ba}{ii} line
compared to a synthesis (solid red) with A(Ba) from the upper limit.
The S/N of 130 is highlighted by the dashed blue lines.}
\label{fig:baul}
\end{figure}

We derived significant upper limits for Na, and the star is found to be poor in Na (see Fig.\,\ref{fig:naul}).
We estimated S/N=85\ in the wavelength range of the strongest \ion{Na}{i} D1 line at 588.9\,nm.
By using Cayrel's formula \citep[see][]{cayrel88}, we derived the minimum equivalent width (EW) detectable.
We multiplied this value by a factor of three  
and derived the abundance by interpolating in a curve of growth.
The observed spectrum compared to the synthesis highlighting the upper limit is shown in Fig.\,\ref{fig:naul}.
We also derived a significant upper limit for Al using the same technique used to determine Na (with S/N=85, see Fig.\,\ref{fig:alul}).
By investigating the 407\,nm \ion{Sr}{ii} line (S/N=90), and proceeding as for Na and Al, we concluded that the star is surely under-abundant in Sr (see Fig.\,\ref{fig:srul}).
With the NLTE correction of 0.14\,dex for Sr derived by \citet{leo12}, the star is still poor in heavy elements, as is frequently the case
in extremely metal-poor stars \citep[see][]{francois07}.
For Ba (S/N=130), the upper limit is less stringent (see Fig.\,\ref{fig:baul})
indicating that the star is compatible with being slightly enhanced in Ba.
The abundances, the upper limits and the NLTE corrections adopted are reported in Table\,\ref{tab:abbo}.

\begin{figure}
\centering
\includegraphics[width=\hsize,clip=true]{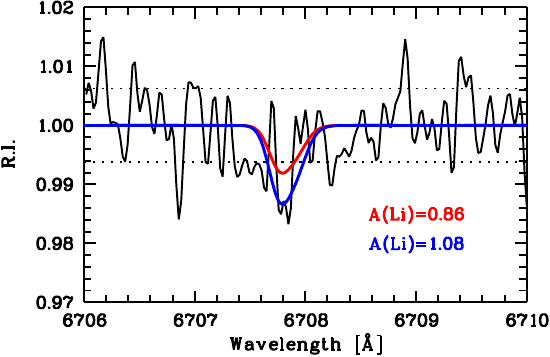}
\caption{Observed spectra (solid black) in the wavelength of the 670.7\,nm \ion{Li}{i} doublet
compared to synthesis (solid red and blue).
The $1\sigma$ S/N is highlighted by the dashed blue lines.}
\label{fig:liul}
\end{figure}

We do not clearly see the \ion{Li}{i} doublet at 670.7\,nm. 
Using Cayrel's formula \citep[see][S/N=150]{cayrel88}, we derived $\rm A(Li)<0.88$ at $3 \sigma$ and $\rm A(Li)<1.08$ at $5 \sigma$.
However, there was a hint of a feature slightly exceeding the noise level.
We smoothed the spectrum to increase the S/N per pixel, but the feature did not appear clearly as the Li doublet, nor did it disappear.
We looked at the spectra individually to see if the hint corresponds to a specific observation, but this feature
builds up by adding the observations.
We conclude that there is a tentative but very uncertain Li detection at $\rm A(Li)\sim 1.08$.
The detection is very uncertain because the feature at 670.8\,nm has a different shape than the synthesis (see Fig.\,\ref{fig:liul} for the comparison of the observed with synthetic spectra).

A fundamental point for this star is the carbon abundance, 
in order to confirm whether or not the star is carbon normal, i.e. not a carbon-enhanced metal-poor (CEMP) star. However, nitrogen is also important to derive.
Neither C from the G-band nor N from the NH band are visible in the observed spectrum.
Since the NH band at 336\,nm is expected to be much weaker than the CH lines in the G-band, and no new data is available at 336\,nm, we concentrated on the determination of an upper limit for C. 
We fit the NH band, by which we mean that we fit the noise in the region with syntheses based on the new stellar parameters, which led to weaker synthetic features. By using the line-list from \citet{fernando18} 
we derived from the fit: $\rm A(N)=4.42\pm 0.15$. We therefore conclude that $\rm A(N)<4.87$, which is the results of the fit added by $3\sigma$, and it is higher than the value in \citet{leo12} because of the higher surface gravity now adopted. 

\subsection{Carbon abundance} \label{sec:carbon}
\begin{figure}
\centering
\includegraphics[width=\hsize,clip=true]{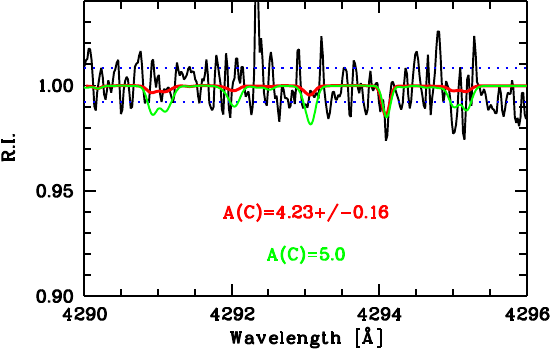}\\
\includegraphics[width=\hsize,clip=true]{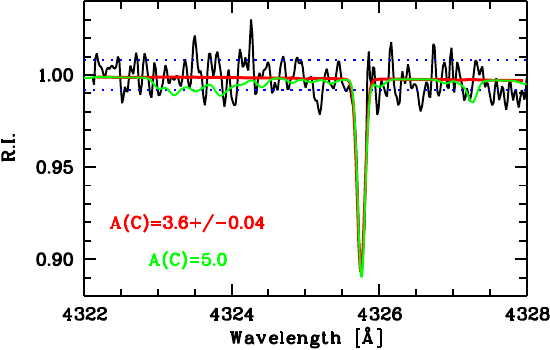}
\caption{Observed spectra (solid black) in the wavelength of the G-band compared to the best fit (solid red) and a synthesis (solid green) to visualise the strongest CH features.
The S/N is highlighted by the dashed blue lines. The visible strong line is \ion{Fe}{i}.}
\label{fig:c}
\end{figure}

We computed synthetic profiles with SYNTHE using the ATLAS\,12 model computed for the star.
We fit the G-band by choosing the wavelength ranges hosting the strongest CH transitions.
The molecular data of the CH lines are from \citet{masseron14}.
We fit a metallic line (a \ion{Fe}{i} line is present in the wavelength range) with the CH features. We adjusted $\rm A{\rm (Fe)}$ in the computation of the synthesis to fit the line as well as possible. 
This was done in order to anchor the shift 
in wavelength to this strong and well-defined line.
We believe that in this spectral range, besides the \ion{Fe}{i} line, there is only noise,
corresponding to $\rm S/N\sim 100$, compatible with a weak G-band.
Fitting ranges in the G-band including an atomic line and the strongest CH features, we derived values in the range:
$\rm 3.60 < A(C) < 4.23$.
In the wavelength range around 432\,nm, from the fit we derived a very low carbon value ($\rm A(C)=3.60$), making the star poor in C, in line with the $\alpha$-elements.
In the wavelength range around 429\,nm and 430\,nm, we find $\rm A(C)=4.23$, consistent with the 
results by \citet{FS6} in a sample of unmixed stars at a metal-poor regime ($\rm -4.2<[Fe/H]<-2$).
In Fig.\,\ref{fig:c}, the fits are shown.
From the fit in the G-band, we deduced that $\rm A(C)<4.71$, which is the higher result from the fit added by $3\sigma$ of its uncertainty,
which implies $\rm [C/H]<-3.79$ and $\rm [C/Fe]<0.94$.

\begin{figure}
\centering
\includegraphics[width=\hsize,clip=true]{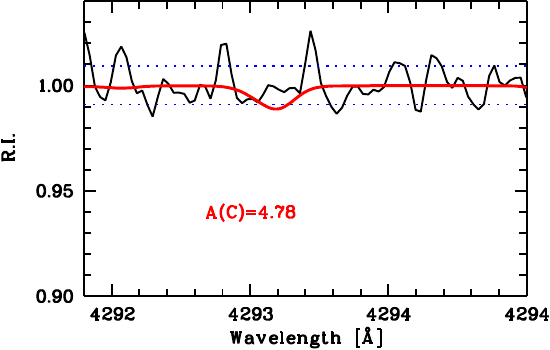}\\
\includegraphics[width=\hsize,clip=true]{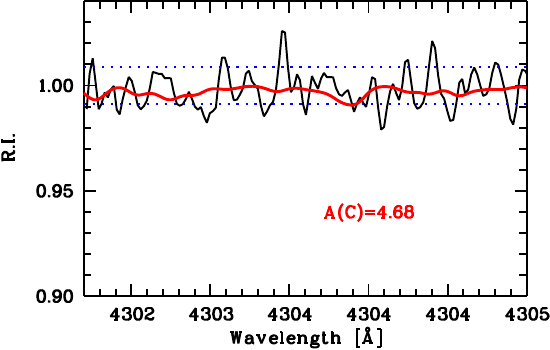}\\
\includegraphics[width=\hsize,clip=true]{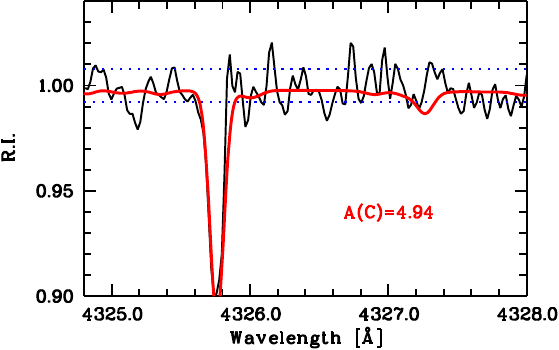}
\caption{Observed spectra (solid black) in the wavelength of the G-band compared to the synthesis with
$3\sigma$ (solid red) abundance in the correspondence of the three strongest feature in the G-band.
The S/N is highlighted by the dashed blue lines.}
\label{fig:cul}
\end{figure}

As for the other elements for which we derived an upper limit, we selected the strongest CH features
in the G-band, computed the EW at $3\sigma$  according to Cayrel's formula and interpolated in a 
computed curve of growth. The results are shown in Fig.\,\ref{fig:cul}, and the upper limit we derived at $3\sigma$ is $\rm A(C)<4.68$, which provides $\rm [C/H]<-3.82$ and $\rm [C/Fe]<0.91$
with the 1D-LTE Fe value from Table\,\ref{tab:abbo}.
If we apply the NLTE correction on Fe, we derive  $\rm [C/Fe]<0.79$.

The upper limits that we derived for carbon are generally lower than the ones provided by \citet{lagae23}. This is due to 1) the improved S/N of our observed spectrum in the region of the G-band, which doubled with the new observations, and 2) a different methodology that we followed: while Lagae and collaborators consider a $\chi^2$ between synthesis and observation computed over a wider wavelength range, we focused on localised strong CH features that have the chance to stand out of the noise.  

\begin{figure}
\centering
\includegraphics[width=\hsize,clip=true]{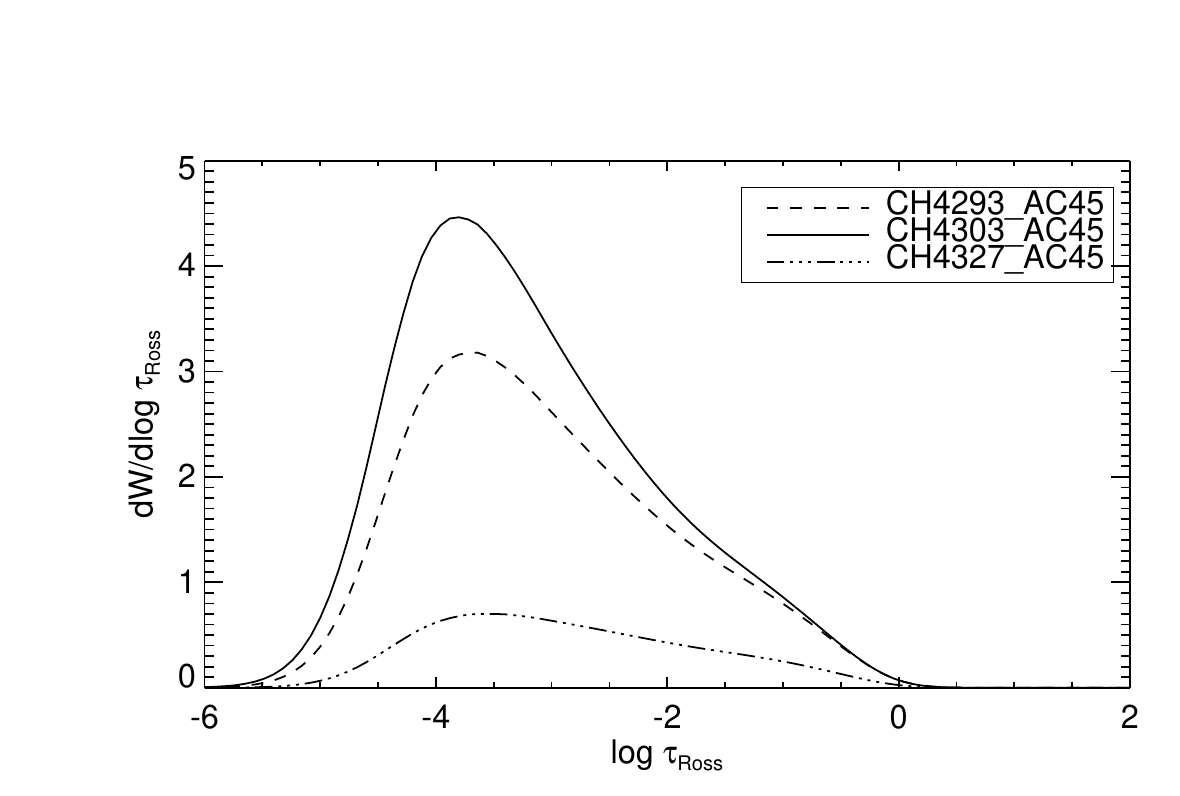}
\caption{EW contribution function, ${\rm d}W/{\rm d}\log \tau_{\rm
    Ross}$ for three of the strongest CH features in the G-band, located at
  $4293$\,\AA\ (dashed), $4303$\,\AA\ (solid), and $4327$\,\AA\ (dash-dotted),
  respectively, as computed from the 3D model shown in Fig.\,\ref{fig:ttau}. 
  The area under each curve gives the line's EW
  in m\AA\ for the disk-integrated stellar spectrum.}
\label{fig:contf}
\end{figure}

The 3D versus 1D temperature differences (see Fig.\,\ref{fig:ttau}) are expected to have consequences for
the spectral line formation and hence for the abundances derived from the two
types of model atmospheres. The strongest CH lines in the G-band, for example, form over a wide range of optical depth, centred on $\log \tau_{\rm Ross}
\approx -4$, as demonstrated by the relevant contribution functions shown in
Fig.\,\ref{fig:contf}. For a given carbon abundance, the 3D models will therefore predict stronger CH lines than the corresponding 1D model.

\begin{figure}
\centering
\includegraphics[width=\hsize,clip=true]{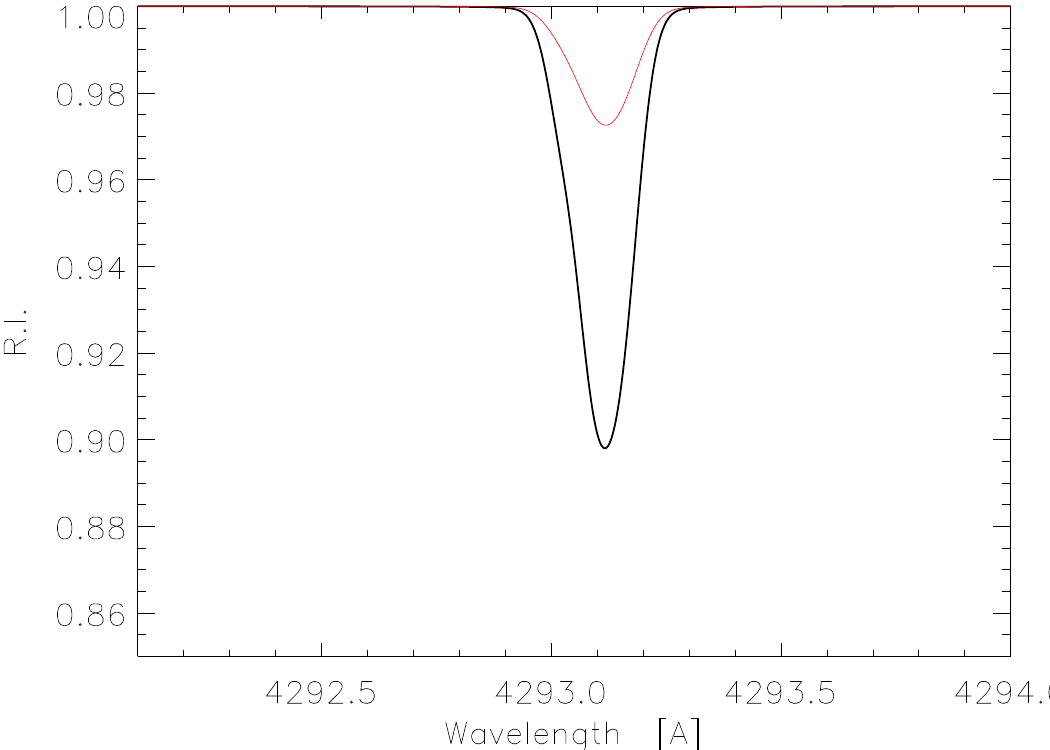}
\caption{Synthetic 3D profile (solid black, EW=1.57\,pm) compared to the 1D synthesis (solid red, EW=0.44\,pm) of the CH feature at 429\,nm in the case of A(C)=4.75. We note that the carbon feature is a blend of several components, such that it appears asymmetric even in the 1D synthesis.}
\label{fig:synt3d1x}
\end{figure}

For a quantitative determination of the 3D effects on the CH lines in the
G-band, we computed synthetic spectral line profiles with the Linfor3D line
formation code \citep{Linfor3D} from (i) the 3D model and (ii) a reference 1D model atmosphere
with identical stellar parameters and using the same opacity tables as
the 3D model. Figure\,\ref{fig:synt3d1x} shows the 3D and 1D synthetic
profiles of the CH feature at $\lambda\,4293$\,\AA, for a given carbon abundance
of A(C)=4.75. The comparison clearly demonstrates that the 3D lines are
significantly stronger than in the 1D case. Quantitatively, the carbon
abundance deduced by means of a 3D analysis of a typical CH feature in the G-band
is about $0.5$\,dex lower than the standard 1D result. This was explored in detail in \citet{Gallagher2016}. The magnitude of
this 3D correction is found to be independent of the spatial resolution of the 
3D model.
From the strongest three CH features used to derive the A(C) upper limit, we derived an average 3D correction of $-0.53$ (with A(C)=4.5).
The value increases with increasing A(C) abundance (it is $-0.72$ in the case of A(C)=5.75).
By applying the 3D correction on the C upper limit, we derived  $\rm [C/Fe]<0.26$.

NLTE calculations for CH have only recently been conducted for a limited number of stellar parameter combinations by \citet{popa23}. They find, in the most extreme case of a metal-poor giant, a NLTE correction of $+0.2$\,dex. We consider this an upper limit of a possible NLTE correction to be present here, since \leo\ is a dwarf where collisions play a more pronounced role, thus driving
the system towards LTE. In addition, \citet{deshmukh23} find negligible deviations from chemical equilibrium of CH in the photospheres of metal-poor turn-off stars. The higher gravity and slightly lower effective temperature of \leo\ suggest that it is unlikely that chemical equilibrium comes out differently in comparison to the situation at the turn-off. Hence, it appears rather unlikely that departures from LTE or chemical equilibrium alter the overall picture described before qualitatively.

\section{Discussion and conclusions}

To derive the upper limit on the metallicity, Z, of the star, we applied the abundances in Table\,\ref{tab:abbo} and the $3\sigma$ upper limits. For the elements for which we have no information, we applied the solar-scaled value, except for N and O for which we applied a value scaled with the C upper limit. We derived $\rm Z<1.915\times 10^{-6}$ ($\rm Z<1.252\times 10^{-4} Z_\odot$ with $\rm Z_\odot$ from \citealt{caffausun} or $\rm Z<1.378\times 10^{-4} Z_\odot$ with $\rm Z_\odot$ from \citealt{asplund21}). This value relies on the 1D-LTE abundances.
The NLTE corrections would increase the metallicity, but only of the elements that represent a minority in the stellar photosphere: an increase by 0.5\,dex in Fe would bring the Z upper limit to $1.970\times 10^{-6}$.
However, if corrections have to be applied, the 3D corrections for C should be taken into account, reducing the stellar metallicity by a larger amount, because carbon is much more abundant in the stellar photosphere than iron.
By applying the NLTE correction from Table\,\ref{tab:abbo} and the 3D correction on C, we derived $\rm Z<1.660\times 10^{-6}$.

With the 1D upper limit $\rm [C/H]<-3.82$ and adopting the scaled value for oxygen, for this star, we derived $\rm D<-3.71$.
By applying the 3D correction on the C abundance and leaving the [O/H] upper limit unchanged, we derived $\rm D<-4.23$.
These values suggest that the star was formed through dust cooling, as suggested by several authors \citep[see][]{schneider12,klessen12,chiaki14,bovino16}, but the fragmentation scenario presented by \citet{greif10} is also a possibility for the formation of this star. 

\leo\ is a peculiar star for several reasons: (i) it is extremely low in metallicity (ii) with no enhancement in carbon; (iii) it is not enhanced in the $\alpha$-elements, and (iv) it is hosted in a Galactic disc orbit.

Few stars with $\rm [Fe/H]<-4.5$\,dex are known, and just two stars \citep[this star and Pristine\,J221.8781+09.7844][]{starkenburg18} do not have a measurable G-band, putting them outside the CEMP class that hosts all the other stars known in this iron-abundance range.
\leo\ is, according to our analysis, the only star with $\rm[Fe/H]<-4.5$\,dex not satisfying the condition $\rm [C/Fe]>1$\,dex proposed by \citet{beers05} to define CEMP stars.
Pristine\,J221.8781+09.7844 has a loose [C/Fe] upper limit \citep[2.3 dex in][with S/N=90 at 400\,nm]{lardo21}, but the G-band is not visible in its spectrum either.
The lower the [Fe/H] ratio, the more difficult is to reach a $\rm [C/Fe]<1$\,dex upper limit, according to the definition of CEMP stars in \citet{beers05}.
Looking at the ultra iron-poor stars in \citet{topos2}, the low carbon band is around $\rm A(C)\sim 6.7$, such that $\rm A(C)< 5.5$ could be a better definition for a star not belonging to the CEMP class in the case that it has $\rm [Fe/H]<-4.5$.

\leo\ shows a feature at the wavelength of the 670.7\,nm Li doublet. This is a tentative Li detection providing $\rm A(Li)=1.08$ (see Table\,\ref{tab:abbo}), a value well below the Spite plateau \citep{spite82}. 
In any case, whether with a low Li abundance or entirely free of Li, the star could have destroyed the Li during its life (e.g. if it were a blue straggler) or have formed from a gas cloud already depleted in Li.
Also, HE\,0107\,5240 \citep{christlieb02} has less lithium than expected from its parameters.
In fact, $\rm A(Li)<0.50$ \citep[see][]{aguado22}, and the star should still have $\rm A(Li)\sim 1$ \citep{mucciarelli22}.
This can be an indication that the ultra Fe-poor stars formed from a cloud depleted in Li.
But, to add complexity to the picture, we recall that at the ultra Fe-poor regime, there are also stars with the Li abundance on the Spite plateau \citep[e.g. SDSS\,J002314.00+030758.0][]{aguado19} or that were on the Spite plateau before diluting
the Li following the first dredge-up \citep[e.g. SMSS\,J031300.36-670839.3][]{keller14}.

The relatively low contents of Al and Na are still compatible with extremely metal-poor (EMP) stars \citep[see][]{cayrel04}. The stringent Sr upper limit puts the star as not rich in heavy elements but it is still consistent with the high-quality EMP sample investigated by \citet{francois07}.
The star is in a pro-grade disc orbit, confined in the Galactic disc. 
Due to the poor metal content of this star, \citet[][see also references therein]{klessen23} suggest that it formed through dust cooling.

The two considerations, that 1) the star is in a Galactic disc orbit and 2) its photospheric metal content is low, may invite to consider \leo\ as a Pop\,III star, polluted by the interstellar gas during its long (probably more than 13\,Gyr) journey through the Galaxy.
\citet{yoshii81} suggested that main-sequence stars accrete surface material through encounters with gas and that the effect is more evident for stars with low-eccentricity orbits, with amount of the accreted material depending on the dimension and obviously on the number of the encountered clouds. The number of encountered clouds is expected to be large due to the long life of this star and the higher number of gas clouds on the Galactic disc than on the Galactic halo;
the relatively low [Mg/Fe] and [Ti/Fe] ratios in \leo\ are consistent with the pollution expected by \citet{johnson15} in a main-sequence Pop\,III star in the case of $\rm \rho_{grain}=3 g cm^{-3}$, but, in this case, the expected [Si/Fe] would be expected to be lower by about 0.3\,dex. 

Our investigation
confirms \leo\ to be the most metal poor object known to date and does not support
any enhancement in carbon.
This leaves open the routes of dust cooling or fragmentation to explain its formation.


\begin{acknowledgements}
We wish to thank Rosine Lallement for providing us the extinction for the star here investigated, and Cis Lagae for discussions on the determination of upper abundance limits. EC and PB acknowledge support from the ERC advanced grant N. 835087 -- SPIAKID. This work has made use of data from the European Space Agency (ESA) mission
{\it Gaia} (\url{https://www.cosmos.esa.int/gaia}), processed by the {\it Gaia}
Data Processing and Analysis Consortium (DPAC,
\url{https://www.cosmos.esa.int/web/gaia/dpac/consortium}). Funding for the DPAC
has been provided by national institutions, in particular the institutions
participating in the {\it Gaia} Multilateral Agreement.
This research has made use of the SIMBAD database, operated at CDS, Strasbourg, France.
\end{acknowledgements}





%


   \bibliographystyle{aa} 

   \bibliography{biblio} 

%


\begin{appendix}

\section{Lines used}

In Table\,\ref{tab:lines}, the atomic lines investigated are listed.\\

\topcaption{\label{tab:lines} Atomic data.}
\tablefirsthead{ \toprule 
\multicolumn{1}{c}{Element}&\multicolumn{1}{r}{$\lambda$}&\multicolumn{1}{r}{\loggf}&\multicolumn{1}{r}{$\rm E_{\rm low}$}\\ 
\multicolumn{1}{c}{ }&\multicolumn{1}{r}{[nm]}&\multicolumn{1}{r}{ }&\multicolumn{1}{r}{$\rm cm^{-1}$}\\ 
}
\tablehead{%
\multicolumn{4}{l}%
{\bfseries \tablename\  \thetable{. Continued.}} \\
\toprule
\multicolumn{1}{c}{Element}&\multicolumn{1}{r}{$\lambda$}&\multicolumn{1}{r}{\loggf}&\multicolumn{1}{r}{$\rm E_{\rm low}$}\\ 
\multicolumn{1}{c}{ }&\multicolumn{1}{r}{[nm]}&\multicolumn{1}{r}{ }&\multicolumn{1}{r}{$\rm cm^{-1}$}\\ 
\midrule}
\tabletail{%
\midrule \multicolumn{4}{r}{{}} \\}
\tablelasttail{%
\\\midrule
}
\begin{supertabular}{lrrr}
\hline
\ion{Na}{i} &  588.9951 & $ 0.108$ &       0.0   \\
\ion{Mg}{i} &  382.9355 & $-0.227$ &   21850.404 \\ 
\ion{Mg}{i} &  383.2299 & $-0.353$ &   21870.465 \\ 
\ion{Mg}{i} &  383.2304 & $ 0.125$ &   21870.465 \\ 
\ion{Mg}{i} &  383.8290 & $-1.527$ &   21911.178 \\ 
\ion{Mg}{i} &  383.8292 & $ 0.397$ &   21911.178 \\ 
\ion{Mg}{i} &  383.8295 & $-0.351$ &   21911.178 \\ 
\ion{Al}{i} &  394.3999 & $-0.635$ &       0.0   \\ 
\ion{Al}{i} &  394.4005 & $-0.635$ &       0.0   \\ 
\ion{Al}{i} &  394.4007 & $-0.635$ &       0.0   \\ 
\ion{Al}{i} &  394.4013 & $-0.635$ &       0.0   \\ 
\ion{Si}{i} &  390.5523 & $-1.041$ &  15394.370  \\ 
\ion{Ca}{i} &  422.6728 & $ 0.244$ &       0.0   \\ 
\ion{Ti}{i} &  336.1212 & $ 0.410$ &     225.704 \\ 
\ion{Fe}{i} &  344.0988 & $-0.958$ &     415.933 \\
\ion{Fe}{i} &  346.5860 & $-1.192$ &     888.132 \\
\ion{Fe}{i} &  347.5450 & $-1.054$ &     704.007 \\
\ion{Fe}{i} &  347.6702 & $-1.507$ &     978.074 \\
\ion{Fe}{i} &  349.0574 & $-1.105$ &     415.933 \\
\ion{Fe}{i} &  355.8515 & $-0.629$ &    7985.785 \\
\ion{Fe}{i} &  356.5379 & $-0.133$ &    7728.060 \\
\ion{Fe}{i} &  357.0097 & $ 0.153$ &    7376.764 \\
\ion{Fe}{i} &  357.0254 & $ 0.650$ &   22650.416 \\
\ion{Fe}{i} &  358.1193 & $ 0.406$ &    6928.268 \\
\ion{Fe}{i} &  358.5319 & $-0.802$ &    7728.060 \\
\ion{Fe}{i} &  360.8859 & $-0.100$ &    8154.714 \\
\ion{Fe}{i} &  361.8768 & $-0.003$ &    7985.785 \\
\ion{Fe}{i} &  363.1463 & $-0.036$ &    7728.060 \\
\ion{Fe}{i} &  367.9913 & $-1.599$ &       0.000 \\
\ion{Fe}{i} &  370.5565 & $-1.334$ &     415.933 \\
\ion{Fe}{i} &  370.9246 & $-0.646$ &    7376.764 \\
\ion{Fe}{i} &  372.7619 & $-0.631$ &    7728.060 \\
\ion{Fe}{i} &  374.5899 & $-1.335$ &     978.074 \\
\ion{Fe}{i} &  375.8233 & $-0.027$ &    7728.060 \\
\ion{Fe}{i} &  376.3789 & $-0.238$ &    7985.785 \\
\ion{Fe}{i} &  376.7192 & $-0.389$ &    8154.714 \\
\ion{Fe}{i} &  378.7880 & $-0.859$ &    8154.714 \\
\ion{Fe}{i} &  381.2964 & $-1.064$ &    7728.060 \\
\ion{Fe}{i} &  381.3058 & $-1.074$ &   20874.482 \\
\ion{Fe}{i} &  381.5840 & $ 0.232$ &   11976.239 \\
\ion{Fe}{i} &  382.0425 & $ 0.119$ &    6928.268 \\
\ion{Fe}{i} &  382.4304 & $-0.033$ &   26627.609 \\
\ion{Fe}{i} &  382.4443 & $-1.362$ &       0.000 \\
\ion{Fe}{i} &  382.5881 & $-0.037$ &    7376.764 \\
\ion{Fe}{i} &  382.7822 & $ 0.062$ &   12560.934 \\
\ion{Fe}{i} &  384.0437 & $-0.506$ &    7985.785 \\
\ion{Fe}{i} &  384.1048 & $-0.045$ &   12968.554 \\
\ion{Fe}{i} &  384.9966 & $-0.871$ &    8154.714 \\
\ion{Fe}{i} &  385.6371 & $-1.286$ &     415.933 \\
\ion{Fe}{i} &  386.5523 & $-0.982$ &    8154.714 \\
\ion{Fe}{i} &  387.2501 & $-0.928$ &    7985.785 \\
\ion{Fe}{i} &  387.8573 & $-1.379$ &     704.007 \\
\ion{Fe}{i} &  387.8671 & $-0.955$ &   19788.252 \\
\ion{Fe}{i} &  389.5656 & $-1.670$ &     888.132 \\
\ion{Fe}{i} &  389.9707 & $-1.531$ &     704.007 \\
\ion{Fe}{i} &  390.2945 & $-0.466$ &   12560.934 \\
\ion{Fe}{i} &  390.6479 & $-2.243$ &     888.132 \\
\ion{Fe}{i} &  392.0257 & $-1.746$ &     978.074 \\
\ion{Fe}{i} &  392.2911 & $-1.651$ &     415.933 \\
\ion{Fe}{i} &  392.7919 & $-1.522$ &     888.132 \\
\ion{Fe}{i} &  393.0296 & $-1.491$ &     704.007 \\
\ion{Fe}{i} &  400.5241 & $-0.610$ &   12560.934 \\
\ion{Fe}{i} &  404.5812 & $ 0.280$ &   11976.239 \\
\ion{Fe}{i} &  406.3594 & $ 0.062$ &   12560.934 \\
\ion{Fe}{i} &  407.1738 & $-0.022$ &   12968.554 \\
\ion{Fe}{i} &  413.2058 & $-0.675$ &   12968.554 \\
\ion{Fe}{i} &  420.2029 & $-0.708$ &   11977.335 \\
\ion{Fe}{i} &  425.0787 & $-0.714$ &   12558.054 \\
\ion{Fe}{i} &  427.1760 & $-0.164$ &   11977.335 \\
\ion{Fe}{i} &  430.7902 & $-0.073$ &   12558.054 \\
\ion{Fe}{i} &  432.5762 & $ 0.006$ &   12969.396 \\
\ion{Fe}{i} &  438.3544 & $ 0.200$ &   11977.335 \\
\ion{Fe}{i} &  440.4750 & $-0.142$ &   12558.054 \\
\ion{Fe}{i} &  441.5122 & $-0.615$ &   12969.396 \\
\ion{Ni}{i} &  345.8460 & $-0.160$ &    1713.087 \\ 
\ion{Ni}{i} &  349.2952 & $-0.216$ &     879.816 \\ 
\ion{Ni}{i} &  349.2954 & $-0.216$ &     879.816 \\ 
\ion{Ni}{i} &  349.2955 & $-0.216$ &     879.816 \\ 
\ion{Ni}{i} &  349.2956 & $-0.216$ &     879.816 \\ 
\ion{Ni}{i} &  349.2957 & $-0.216$ &     879.816 \\ 
\ion{Ni}{i} &  352.4531 & $ 0.044$ &     204.787 \\ 
\ion{Ni}{i} &  352.4534 & $ 0.044$ &     204.787 \\ 
\ion{Ni}{i} &  352.4535 & $ 0.044$ &     204.787 \\ 
\ion{Ni}{i} &  352.4536 & $ 0.044$ &     204.787 \\ 
\ion{Ni}{i} &  352.4537 & $ 0.044$ &     204.787 \\ 
\ion{Ni}{i} &  361.9389 & $ 0.045$ &    3409.937 \\ 
\ion{Ni}{i} &  361.9390 & $ 0.045$ &    3409.937 \\ 
\ion{Ni}{i} &  361.9391 & $ 0.045$ &    3409.937 \\ 
\ion{Ni}{i} &  385.8297 & $-0.865$ &    3409.937 \\ 
\ion{Sr}{ii}&  407.7709 & $ 0.148$ &       0.0   \\
\ion{Ba}{ii}&  455.4029 & $ 0.170$ &       0.0   \\
\hline
\end{supertabular}

\end{appendix}

\end{document}